\documentclass[prb,preprint,groupedaddress,showpacs]{revtex4-1}

\usepackage{graphicx}
\usepackage{amsmath}
\usepackage{epstopdf}
\begin{document}

%\preprint{}

\title{Extended Malus' Law with THz metallic metamaterials for sensitive detection with giant tunable quality factor.}


\author{Xavier Romain$^{1,*}$, Fadi Baida$^{1}$ and Philippe Boyer$^1$}
\email[]{xavier.romain@femto-st.fr}
%\thanks{}
\affiliation{$^1$Institut FEMTO-ST, UMR 6174 CNRS, D\' epartement d\'\ optique P. M. Duffieux,
Universit\' e de Bourgogne Franche-Comt\' e, 25030 Besan\c con Cedex, France}


\date{\today}

\begin{abstract}
We study a polarizer-analyzer mounting for the terahertz regime with perfectly conducting metallic polarizers made of a periodic subwavelength pattern. We analytically investigate the influence on the transmission response of the multiple reflections  which occur between polarizer and analyzer with a renewed Jones formalism. We demonstrate that this interaction leads to a modified transmission response: the extended Malus' Law. In addition, we show that the transmission response can be controlled by the distance between polarizer and analyzer. For particular set-ups, the mounting exhibits extremely sensitive transmission responses. This interesting feature can be employed for high precision sensing and characterization applications. We specifically propose a general design for measuring electro-optical response of materials in the terahertz domain allowing detection of refractive index variations as small as $10^{-5}$.
\end{abstract}

\pacs{07.57.Kp, 07.60.Fs, 42.25.Ja, 42.79.Ci, 42.79.Qx, 78.20.Bh, 78.67.Pt ,81.05.Xj}
%\keywords{}

\maketitle

%\section{}
%\subsection{}
%\subsubsection{}

\section{Introduction}

Unusual light phenomena can be observed and engineered when using subwavelength patterned materials also known as metamaterials \cite{veselago_electrodynamics_1968,pendry_negative_2000}. Since the advent of metamaterials, the extraordinary optical transmission has been one of the extensively studied phenomenon this last decade \cite{ebbesen_extraordinary_1998,baida_light_2002,moreau_light_2003,baida_origin_2004,salvi_annular_2005,poujet_90_2007,baida_enhanced_2010}. Nowadays, extraordinary optical transmission may be used for polarization applications such as anisotropic plates \cite{baida_enhanced-transmission_2011,dahdah_nanocoaxial_2012} and polarization manipulation \cite{grady_terahertz_2013,shen_ultra-high-efficiency_2014,pfeiffer_controlling_2014} with higher performances than conventional components used in visible/IR spectral domain. For the terahertz (THz) domain where natural materials don't basically exhibit efficient dichroism property, it is now well-known that linear polarizers are commonly obtained with frequency selective surfaces or with metallic gratings. Some papers proposed to demonstrate the polarizing properties of periodic subwavelength apertures with the use of the well known Malus' Law \cite{gordon_strong_2004,degiron_optical_2004,sarrazin_polarization_2004,dimaio_polarization-dependent_2006}. However, recent experimental results are clearly in contradiction with the output transmission predicted theoretically by this law when using subwavelength patterned metallic polarizers \cite{huang_break_2014,Zhang:15}. One explanation given in \cite{Zhang:15} for this breakthrough involves the reflections between the plates.

\begin{figure}
\centering\includegraphics[width=\textwidth]{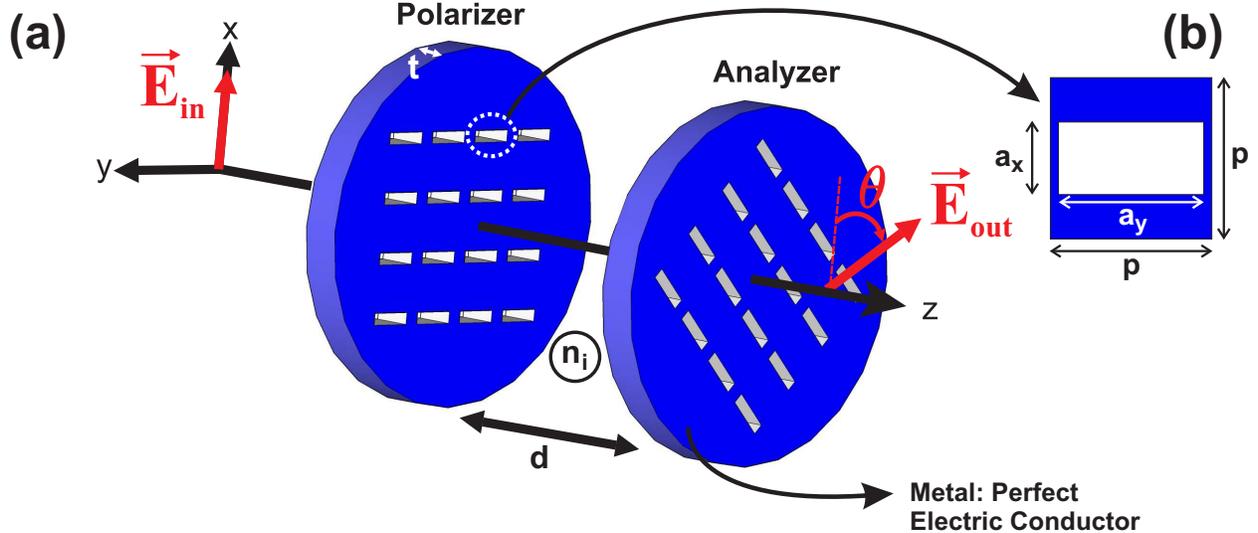}
\caption{(a) 3D view of the Polarizer-Analyzer Mounting (PAM) where $d$ is the distance between polarizer and analyzer, $n_i$ is the refractive index of the inner space between the two metallic plates, $t$ is the plate thickness and $\theta$ is the angular difference between polarizer and analyzer axes. (b) 2D view of the considered subwavelength pattern with $p$, the bi-periodicity, $a_x$ and $a_y$ the rectangle's width and length respectively.}
\label{schPAM}
\end{figure}

In this paper, we theoretically report an extended Malus' Law for metallic Polarizer-Analyzer Mounting (PAM). For a specific configuration based on multiple reflections inside the PAM, we propose one principle of new ultrasensitive sensors with giant quality factors controlled by the angle between the polarizer axes. Precisely, we study a PAM (as illustrated in fig. \ref{schPAM}) made of parallel metallic polarizers with biperiodic subwavelength grating where each periodic cell consists in a single rectangular aperture. The angle between polarizer and analyzer axes is denoted $\theta. $The periods along $x$- and $y$-axes (at $\theta=0^{\circ}$) are identical and noted $p$. Each rectangular aperture only supports the fundamental $TE_{01}$ guided mode. For one polarizer, the output linear electric polarization is thus defined along the rectangle width axis for the working wavelength ranges chosen as $\lambda>\lambda_{c,TE_{01}}$ where $\lambda_{c,TE_{01}} = 2a_y >p$ (subwavelength approximation), $\lambda_{c,TE_{01}}$ being the cut-off wavelength of $TE_{01}$ fundamental guided mode. The higher order modes $TE_{10}$, $TE_{02}$ and above are evanescent ($\lambda_{c,TE_{10}}$ and $ \lambda_{c,TE_{02}} < p$).  Metal is assumed to be a perfect electric conductor at THz frequencies. 

The analytical extended Malus' law is deduced from a renewed Jones formalism for metallic polarizers \cite{boyer_jones_2014,romain2015extended} based on a monomode modal method \cite{lalanne_one-mode_2000,martin-moreno_theory_2001,baida_enhanced-transmission_2011,boyer_analytical_2012}. We show that these extended Malus' Law basically takes the following form:
\begin{equation}
I_{out}=\vec{E}_{out}\cdot\vec{E}_{out}^*=I_{in}\left|\alpha(\theta,\lambda,L)\right|^2\cos^2\theta
\label{EML}
\end{equation}
where $*$ denotes the complex conjugate, $I_{out}$ is the output electric intensity with $\vec{E}_{out}$ the transmitted electric field, $I_{in}=\vec{E}_{in}\cdot\vec{E}_{in}^*$ is the electric intensity incident on the polarizer. The modulation factor $\alpha$ will be analytically expressed in section 2. Nevertheless, we highlight its dependencies on three main parameters which affect the resonance properties of the studied PAM. First, the $\theta$ dependency causes the substantial discrepancies with the well-known and classical Malus' Law (electric intensity proportional to square cosine of $\theta$). Besides, $\theta$ controls the quality factor of PAM resonances. Second, the coefficient $\alpha$ is an Airy-like spectrally resonant term ($\lambda$ dependency) which ensures a perfect transmission at polarizers resonances \cite{boyer_jones_2014}. Third, we have specified in eq. (\ref{EML}), the dependency of $\alpha$ on the optical path $L=n_id$ where $n_i$ is the refractive index of the inner space between polarizer and analyzer separated by a distance $d$ (see fig. \ref{schPAM}). We will show that this parameter $L$ is linked to multiple reflections between polarizers and controls the sensitivity of PAM resonances. 

In section 2, we present the theoretical formalism  which allows us to derive the extended Malus Law given in eq. (\ref{EML}). To underline the influence of the multiple reflections inside the PAM, we compare it with the one obtained with dichroic polarizers. Afterwards, we numerically investigate the PAM's transmission response to highlight the properties of the extended Malus' Law. Particularly, we show that high sensitivity can be obtained. In the section 3, we take benefits of this interesting property to propose a device combining a good sensitivity, a tunable quality factor, and a high extinction ratio over spectral broad band.

\section{Tunable Transmision Response of a metallic Polarizer-Analyzer  Mounting}
\subsection{Theoretical Framework}
\label{sec:theo}
Our model is based on a monomode modal method \cite{baida_enhanced-transmission_2011,lalanne_one-mode_2000,martin-moreno_theory_2001,boyer_analytical_2012} which consists to consider that only the fundamental guided mode  of the rectangular apertures is excited (propagation along the metal film thickness). It has to be noted that the formalism is also applicable to other common 2D shapes (for example split-ring resonators \cite{smith_composite_2000} and annular apertures \cite{baida_three-dimensional_2003}) and 1D shapes such as wire grids \cite{Yamada:09}. We consider our PAM illuminated at normal incidence. We assume that working wavelengths are higher than the first Rayleigh wavelength which means that only the $0^{th}$ diffracted order in Fourier-Rayleigh expansions is propagative in homogeneous regions inside and outside the PAM. We also consider a far-field approximation in the sense that the evanescent waves are not taken into account in the description of the electromagnetic fields (Fourier-Rayleigh expansions reduced to the single propagative $0^{th}$ diffracted order). This last assumptions is especially verified if $L > \lambda/2$. Nonetheless, evanescent diffracted orders are taken into account for the computation of the transmission and reflection Jones matrices $J_k^{T,R}$ of the $k^{th}$ polarizer ($k \in {1,2}$) that can be expressed as follows \cite{boyer_jones_2014}:
\begin{equation}
J_k^{T,R}=\alpha_{T,R}(\lambda)J_k-\xi_{T,R}I_d
\label{eq:JkTR}
\end{equation}
where $I_d$ is the identity matrix, $\xi_T=0$ and $\xi_R=1$. The terms $\alpha_{T,R}(\lambda)$ are Fabry-Perot-like spectral resonant transmission/reflection complex coefficients for the $k^{th}$ polarizer (readers may find their expressions in eqs. (4) and (5) of ref. \cite{boyer_jones_2014}). The matrix $J_1$ is the conventional transmission Jones matrix of a linear polarizer oriented along the x-axis and $J_2$ is the one of the analyzer that is rotated by an angle $\theta$ counted from the x-axis. Knowing that $J_k^{T,R}$ identifies to the propagative $0^{th}$ diffracted order 2$\times$2 sub-blocks of the full scattering matrix of each polarizers, the scattering propagation algorithm \cite{li_formulation_1996} is used to analytically compute the transmission Jones matrix $J^T_{PAM}$ of the PAM. After tedious calculations, we obtain:
\begin{equation}
J_{PAM}^T=\alpha(\theta,\lambda,L) J_2J_1
\label{JTPAM}
\end{equation}
We focus our attention on the transmitted output intensity and the expression of the reflection Jones matrix of the PAM is not given in this paper. The extended Malus' Law given in eq. (\ref{EML}) is directly derived from the eq. (\ref{JTPAM}), where the term $\alpha$ is expressed by:
\begin{equation}
\alpha(\theta,\lambda,L) = \frac{\alpha_T^2(\lambda) u}{\gamma-u^2\left[1-\alpha_R(\lambda)\right]^2}
\label{eq:alpha}
\end{equation}
with
\begin{equation}
\gamma=\frac{1-u^2\left[1- \alpha_R^2(\lambda) \sin^2\theta\right]}{1-u^2}
\label{eq:gamma}
\end{equation}
where $u = \exp(ik_0L)$ is the propagation term between polarizer and analyzer, with $k_0 = 2\pi /\lambda$. It is important to notice that in our study the extended Malus' Law is evaluated at spectral resonances of $\alpha$  (maxima of $\left|\alpha\right|$). The $\theta$ dependency of $\alpha$ clearly appears in the expression of $\gamma$ ($\sin^2 \theta$ in eq. (\ref{eq:gamma})). This dependency is multiplied by $\alpha^2_R(\lambda)$ which implies that the multiple reflections occurring between the two polarizers are directly linked to the transmission response and provoke the important discrepancies with the classical Malus' Law. Moreover, the term $u^2$ in the numerator of $\gamma$ means that the optical path $L$ controls the influence of multiple reflections on the extended Malus'law variation. 

It has to be noted that a device with similar polarization properties called Malus Fabry-Perot interferometer was theoretically investigated in 1999 by Vallet \textit{et al.} \cite{vallet_malus_1999}. This device consisted in a Fabry-Perot interferometer included inside a PAM (without spectral resonances of polarizing plates) made of crossed polarization beamsplitters. The two mirrors of that device provoke similar multiple reflections to the ones generated by the metallic polarizers in our structure. However, the behaviors of these two kinds of polarizing resonators are different. For the Malus Fabry-Perot interferometer of ref. \cite{vallet_malus_1999}, all polarized transmitted and reflected waves by each polarizer are reflected by mirrors, which makes Fabry-Perot resonances and polarization effects independent. For our structure, the metallic plates play the role of both polarizers and mirrors because, at resonance, only the waves polarized along the rectangle's length axis are reflected inside the PAM. The Fabry-Perot-like resonances between polarizers, and polarization effects are thus closely linked, and this is exploited to perform the efficient application proposed below. 

Our formalism also allows us to compute the classical Malus' Law obtained with dichroic polarizing plates. For the sake of completeness, we first give the general expression of the reflection Jones matrix of the $k^{th}$ polarizer oriented along $x$-axis:
\begin{equation}
J_k^R = \left (\begin{array}{cc}
\alpha_R - 1 & 0 \\
0 & \beta
\end{array}\right )
\label{JkRAbs}
\end{equation}
$\beta$ is the reflection coefficient of one polarizer along the rectangle length axis calculated in accordance with the absorption along this axis only. For metallic polarizers, $\beta = -1$ (no absorption), the eq.(\ref{JkRAbs}) is identical to the eq. (\ref{eq:JkTR}) for $k=1$. For dichroic polarizer, $\beta=0$ (total absorption along the rectangle length axis) meaning that multiple reflections in the PAM are reduced to the ones oriented along the rectangle width axis (term $\alpha_R-1$ in $J_k^R$). These reflections are weak for most of natural dichroic plates. This leads to the expression $\alpha_d$ for $\alpha$ of the eq. (\ref{EML}) in the case of the classical Malus'law when the multiple reflections are not neglected ($\beta=0$ and $\alpha_R\approx 1$ with $\alpha_R\neq 1$):
\begin{equation}
\alpha_d(\theta,\lambda,L) = \frac{\alpha_T^2(\lambda) u}{1 - u^2\left[1-\alpha_R(\lambda)\right]^2\cos^2\theta} \ \ \underset{ \ \ \alpha_R \to 1}{\longrightarrow} \alpha_T^2(\lambda)u
\label{dichroism}
\end{equation}
This equation highlights the discrepancies between the factor $\alpha_d$ found for commonly used dichroic polarizers and the modulation factor $\alpha$ previously obtained for metallic polarizer (eqs. \eqref{eq:alpha} and \eqref{eq:gamma}). We see that $\alpha_d$ is \emph{a priori} dependent on $\theta$ but the factor $u^2\left[1-\alpha_R(\lambda)\right]^2$ relating to multiple reflections may be neglected for the special case of highly absorbing dichroic polarizers ($\alpha_R\to 1$). On the opposite,  the term $u^2\left[1- \alpha_R^2(\lambda) \sin^2\theta\right]$ cannot be neglected in the expression (\ref{eq:gamma}) of $\gamma$ for the case of metallic polarizers. Indeed, we know that $\alpha_R\approx 1$ when $\left|\alpha_T\right|=1$ due to energy balance criterion: $\left |\alpha_T^2 - (\alpha_R - 1)^2\right |= 1$ (see fig. 3 of \cite{boyer_jones_2014}). Consequently and contrary to the extended Malus'law for metallic polarizers, we can suppose that $\alpha_d$ is independent on $\theta$ as it is well-known for dicroic polarizers which leads to $\alpha_T^2(\lambda)u$. The Malus' law takes the form of the classical one which corresponds to  single pass propagation through the PAM.

\subsection{Numerical Results}
We propose a numerical investigation of the PAM's transmission depicted in fig. \ref{schPAM}. We focus on the influence of the optical path $L$. The dimensions of the polarizer and analyzer rectangle are: $a_y/p = 0.9$ and $a_x/p = 0.45$. These values are chosen such as the radiative losses of the metallic polarizer's apertures are maximized (spectrally broadband transmission). Besides, the $a_x/p$ value is set to ensure that only the fundamental mode can propagate in apertures at wavelengths located above the Rayleigh anomaly (monomode regime). In other words, the cut-off wavelength of the second cavity mode is smaller than the first Rayleigh wavelength. The thickness of the polarizer and analyzer is set to $t/p = 1$. In this section, we consider that all the homogeneous regions (including the apertures) are filled with air. For all results, we compute the normalized transmission coefficient: $I_{out}/I_{in}$.

\begin{figure}
\centering\includegraphics[width=\textwidth]{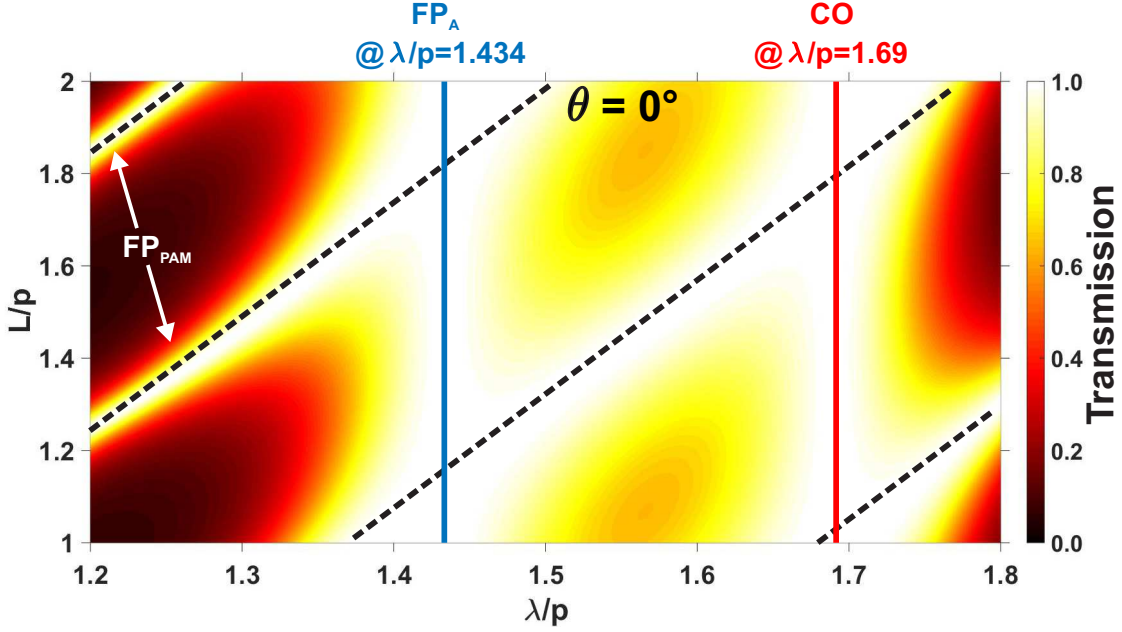}
\caption{Normalized transmitted electric intensity spectra of the PAM versus $L/p$ for $\theta = 0^o$. The parameters are: $a_x/p = 0.45$, $a_y/p = 0.9$ and $t/p = 1.0$. Vertical lines show resonances of $|\alpha|$ ($I_{out}=1$): the one at $\lambda/p = 1.434$ is related to the first harmonic of the Fabry-Perot resonance of the  fundamental mode guided inside the rectangular apertures ($FP_A$) and the other one at $\lambda/p = 1.69$ corresponds to the cut-off of the same mode ($CO$). The $FP_{PAM}$ branches (oblique dashed lines) denotes Fabry-Perot interferences located between polarizer and analyzer.}
\label{IvsLambGap}
\end{figure}

\begin{figure*}
\includegraphics[width=\textwidth]{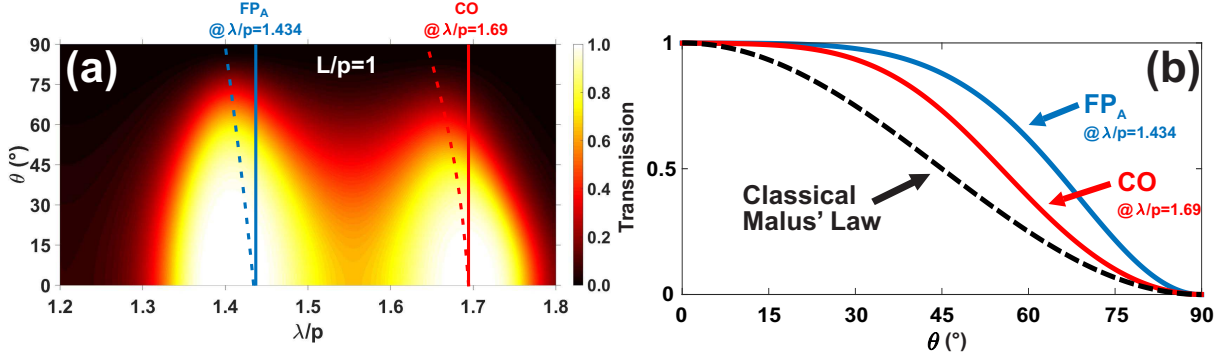}
\caption{\textbf{(a)} Normalized transmission spectra versus $\theta$ for $L/p=1$ (see fig. \ref{IvsLambGap} for other parameters). The curved dashed lines represent the trajectories of the resonance of $\alpha$ ($|\alpha| = 1$). \textbf{(b)} Normalized transmission computed for fixed values of $\lambda/p$ (blue and red vertical solid lines in (a)) and compared with the classical one (dashed black line).}
\label{fig:MalusLawvsLambda}
\end{figure*}

We first calculate the transmitted electric intensity spectrum as a function of the distance $L/p$ for $\theta = 0^o$ ($I_{out}=\left|\alpha\right|^2$) as shown in fig \ref{IvsLambGap} in order to reveal all resonances supported by the whole structure (spectral resonances of $\alpha$). The resonance at $\lambda/p = 1.434$ (blue vertical line) corresponds to the first harmonic of the Fabry-Perot resonance ($FP_A$) of the fundamental mode guided inside the rectangular apertures along the metal thickness. The resonance at $\lambda/p = 1.69$ (red vertical line) corresponds to the cut-off of the same mode ($CO$). Both resonances correspond to transmission resonances of one metallic polarizer: $\left|\alpha_T(\lambda)\right|=1$. The different oblique branches ($FP_{PAM}$ in oblique dashed line) correspond to the Fabry-Perot resonances resulting from the multiple reflections between the two polarizers. The $FP_A$ and $CO$ resonances ensure a high transmission for any value of $L/p$. Thereafter, we will restrict our analysis to the transmission at $FP_A$ and $CO$ resonances related to each polarizer.

Figure \ref{fig:MalusLawvsLambda} (a) shows the transmission spectra as a function of $\theta$ for $L/p=1$. It reveals that resonance wavelength values are affected by the variation of $\theta$ (curved dashed colored lines). Figure \ref{fig:MalusLawvsLambda} (b) shows the transmission at the wavelengths corresponding to vertical colored solid lines (plotted at resonance wavelengths for $\theta = 0^{\circ}$ in fig. \ref{fig:MalusLawvsLambda} (a)) and compared with the classical Malus' Law (dashed black line) for which the Half-Width Half Maximum ($HWHM$) is equal to $\pi/4$. The observed discrepancies confirms the important contribution of the multiple reflections between polarizer and analyzer.

\begin{figure*}
\centering\includegraphics[width=\textwidth]{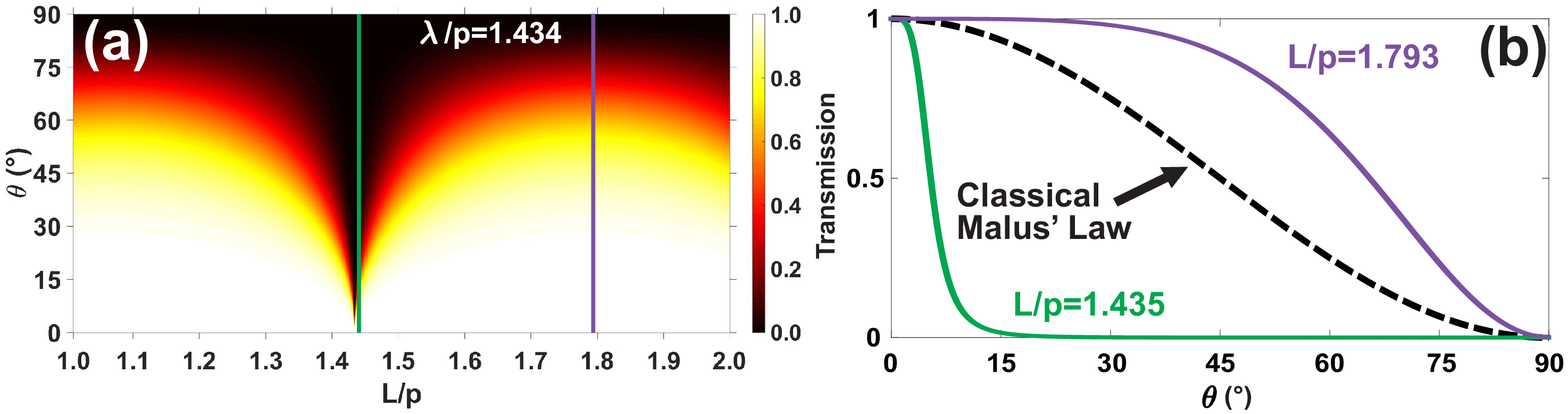}
\caption{\textbf{(a)} Normalized transmitted intensity versus $L/p$ and $\theta$ for $\lambda/p = 1.434$ (see fig. \ref{IvsLambGap} for other parameters). \textbf{(b)} Normalized transmission computed for fixed values of $L/p$ (green and purple vertical dashed lines in (a)) by comparison with the classical one (dashed black line).}
\label{fig:MalusLawvsGap}
\end{figure*}

In fig. \ref{fig:MalusLawvsGap} (a), we choose $\lambda/p = 1.434$ ($FP_A$-resonance) and we plot the transmission as a function of $\theta$ and $L/p$. As mentioned in section \ref{sec:theo}, the optical path $L/p$ is an important parameter that will allow us to tune the PAM transmission. We distinguish two contrasting cases:
\begin{enumerate}
\item When $u^2=1$ which is equivalent to $L= m\lambda/2$ with $m$ a natural integer. We observe an infinitely narrow angle Malus' Law ($HWHM \ll \pi/4$). Precisely, the transmitted electric intensity drops to 0 for this particular value of $L/p$ when $\theta \neq 0^{\circ}$. Indeed, this is explained by the fact that the term $\gamma$ in eq. (\ref{eq:gamma}) diverges when $\theta \neq 0^o$. For $\theta=0^o$, we clearly see that $\gamma=1$ which implies that $\alpha=\alpha_T^2(\lambda)/\alpha_R^2(\lambda)$ approximately equals to $1$ at resonances of $\alpha_T$ ($I_{out}\approx 1$). For the sake of clarity, the Malus'law is shown in the inset of the fig. \ref{fig:MalusLawvsGap} at $L/p=1.435$ (green line) and not exactly at $L/p=1.434$ ($u^2=1$ for $m=2$) for which the transmission results in a Kronecker function:
\begin{equation}
\alpha(\theta)=\delta_{\theta,0}
\end{equation}
\item When $u^2=-1$, which is equivalent to $L=\lambda/4+m'\lambda/2$ with $m'$ a natural integer, we observe broad angle Malus' Law. Precisely, the transmitted electric intensity is approximately constant and remains maximum for a wide range of $\theta$. The transmission versus $\theta$ is shown in the fig. \ref{fig:MalusLawvsGap} (b) (purple line) at $L/p=1.793$ ($u^2=-1$ for $m'=2$). Such a transmission may be seen as a complementary Airy-like function ($HWHM > \pi/4$) with a unity value plateau for small $\theta$. The following equation gives the simple expression of $\alpha$ for the purple line in fig. \ref{fig:MalusLawvsGap} (b) assuming that $\alpha_R=1$ (the computed value being exactly equal to $1.0077+i0.1307$):
\begin{equation}
\alpha(\theta,\lambda)=i(-1)^{m '}\frac{\alpha_T^2(\lambda)}{1-\frac{1}{2}\sin^2\theta}
\end{equation}
\end{enumerate}
Consequently, both narrow and broad angle Malus' Law can be achieved by tuning $L$.

%Knowing that $\sin\theta=\theta+o\left(\theta^2\right)$ for relatively small values of $\theta$, we have $\alpha=-\alpha_T^2/{\left[\cos\theta+o\left(\theta^4\right)\right]}% and so $I_{out}\approx \left|\alpha_T\right|^4$ equal to $1$ at the resonance of $\alpha_T$.

\section{Application to design ultrasensistive THz sensing with giant and tunable Q}
\begin{figure*}
\centering\includegraphics[width=\textwidth]{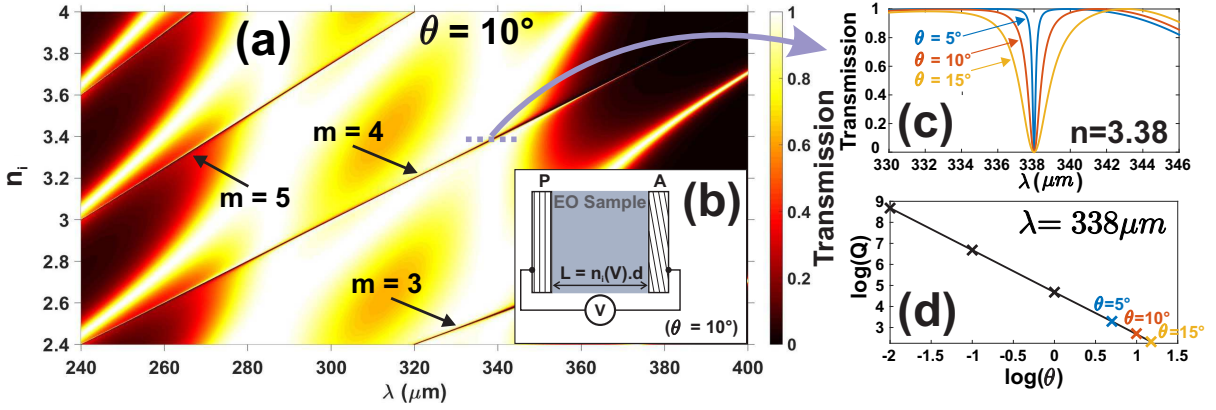}
\caption{\textbf{(a)} Normalized transmitted intensity spectrum vs the refractive index $n_i$, for $\theta = 10^{\circ}$. The parameters are $p=100\mu m$, $a_x = 45\mu m$, $a_y = 90\mu m$, $t=100\mu m$ and $d=100\mu m$. The dark branches correspond to narrow transmission dips when $L = m\lambda/2$. \textbf{(b)} Scheme of the general principle to characterize electro-optical material responses. \textbf{P}: Polarizer (first electrode), \textbf{A}: Analyzer (second electrode), \textbf{V}: applied DC voltage, \textbf{d}: EO material thickness and \textbf{$n_i(V)$}: refractive index of the electro-optical sample. \textbf{(c)} Normalized transmission dips for different values of $\theta$ in degrees, at $n_i = 3.38$. \textbf{(d)} Quality factor Q as a function of $\theta$ at $n_i=3.38$.}
\label{fig:MalusLawvsOPL_theta10}
\end{figure*}

Taking advantage of an infinitely narrow angle Malus' law shown in the previous section for particular values of $L$, we give here the principle of a spectrally sensitive system in the THz domain and with a tunable quality factor. Such a system can be used for many applications as for temperature or pressure sensors, or characterization of an electro-optical (EO) material. We now assume that the middle region sandwiched in-between polarizers with $\theta\neq 0^o$ is filled with an isotropic, homogeneous and transparent EO material (see fig. \ref{fig:MalusLawvsOPL_theta10} (b)). It is interesting to remark that the two metallic polarizers play also the role of electrodes to tune the refractive index $n_i(V)$ of the EO material. For this study, the value of the distance $d$, corresponding to the EO material's thickness, is fixed so that $L$ only varies with its refractive index $n_i$. The dimensions are $p=200\mu m, a_x=90\mu m, a_y=180\mu m$, $t=200\mu m$ and $d=200\mu m $.

Figure \ref{fig:MalusLawvsOPL_theta10} (a) shows the transmitted electric intensity spectrum according to $n_i$. Contrary to the intensity spectra plotted in fig. \ref{IvsLambGap} ($\theta = 0^{\circ}$), oblique and very narrow dark branches appear (when $\theta \neq 0^{\circ}$) in transmission bands corresponding to transmission dips satisfying $n_i=m\lambda/(2d)$ ($u^2=1$). In order to match the refractive index range of most of the material in the THz domain ($n_i$ approximately between $3$ and $4$ \cite{wu_design_1996}), we must consider the branch $m=4$. As a remark, we note that the device can be adapted to any range of refractive index values by adjusting the value of the order $m$. The sensitivity $S$ associated to those dark branches is in our case:
\begin{equation}
S = \frac{\Delta\lambda}{\Delta n_i} = \frac{2d}{m}, \ \ \forall\theta \neq 0^{\circ}
\label{Sens}
\end{equation}
For $m=4$, we have $S = 100\mu m$/RIU. We point out that it is possible to improve the sensitivity by increasing the thickness $d$ (compromise between compactness and sensitivity). 

The figure \ref{fig:MalusLawvsOPL_theta10} (c) shows intensity spectra for $\lambda$ close to $338 \mu m$ ($\lambda/p = 1.69$) for different value of $\theta$ at $n_i=3.38$ which is close to the Gallium phosphide (GaP) refractive index in the THz domain ($n_i \simeq n_{_{GaP}}=3.34$, see \cite{wu_design_1996}). We mention that those dips have high extinction ratios in transmission bands. We also precise that $\theta$ does not affect the sensitivity. However, the width of the transmission dips (or quality factor) can be controlled by adjusting $\theta$, in accordance with results presented in fig. \ref{fig:MalusLawvsGap}. Fig. \ref{fig:MalusLawvsOPL_theta10} (d) shows the variation of the quality factor $Q$ versus $\theta$ (for $n_i=3.38$ at $\lambda=338\mu m$). We see that the quality factor theoretically diverges when $\theta$ tends to $0^{\circ}$ and the linearity of the curve allows us to write:
\begin{equation}
Q = \frac{mA}{\theta^B}, \ \ \forall\theta \neq 0^{\circ}
\label{Qual}
\end{equation}
with $A$ and $B$ two empirical and positive parameters, and $\theta$ expressed in degree. From fig. \ref{fig:MalusLawvsOPL_theta10} (d), we deduce $B=2$ and $A=1.25 \times 10^4$ degrees$^2$ for $m=4$.

%%We have numerically found that $A$ is proportional to $m$ while $B$ does not vary with $m$.

Finally, we are interested in finding a suitable value of $\theta$ to obtain a quality factor matching the resolution of Thz spectrometers (under Rayleigh criterion). Then, we deduce the minimum variation of the refractive index $(\Delta n_i)_{min}$ which can be detected by the device. Heterodyne detectors in THz domain offers spectral resolution ($R=\Delta\lambda/\lambda$) equal to $3.3 \times 10^{-6}$ (see \cite{chusseau2008optoelectronique}). Thus, according to eq. \eqref{Qual}, to reach $Q=1/R =3 \times 10^5$, $\theta$ must be equal to $0.4^{\circ}$ at $n_i=3.38$ and $\lambda=338\mu m$. With such a quality factor, we derive from eq. \eqref{Sens} that $(\Delta n_i)_{min}= \lambda/(S.Q) = 1.1 \times 10^{-5}$.

Consequently, we have designed a very efficient system for THz applications ($S=100\mu m/RIU$, $Q=3.10^5$). By comparison, Ranjan Singh \emph{et al.} \cite{singh_ultrasensitive_2014} has experimentally proposed a metasurface reaching $S=57\mu m$/RIU and $Q=28$. We have to remind that our numerical results are obtained from a theory which assumes a perfect (rigorously identical apertures) and infinite periodicity of the metallic polarizers in addition to a perfect parallelism between polarizer and analyzer. We also assume an isotropic and lossless EO material. We expect that breaking these assumptions may affect the performances of the proposed system.

\section{Conclusion}
In summary, we have given an analytical formalism of an extended Malus' Law with metallic polarizer for the terahertz regime. Our theoretical investigation highlights the important discrepancies with the classical Malus' Law due to the $\theta$ dependency of the modulation factor as well as the multiple reflections inside the PAM which are tunable through the optical path $L$. Indeed, for specific values of $L$ one can obtain broad angle or narrow angle Malus' Law. Then, we designed a structure for characterizing, with high sensitivity and high quality factor, the electro-optical response of terahertz EO material based on an extremely narrow angle Malus' Law. This analytical model of a two-layers stack of subwavelength structures provides new theoretical insights into the interactions between polarizing metamaterials. This simple structure can be seen as the basic component for multi-layered and more complex structures. In future works, we will further use our analytical model as a platform to propose other applications such as high efficiency polarization conversion, high-Q filtering and ultra-sensitive polarimetry.

\begin{acknowledgments}
This project has been performed in cooperation with the Labex ACTION program (contract ANR-11-LABX-0001-01).
\end{acknowledgments}

\bibliography{MLA110116}

\end{document}